\documentclass[epj]{svjour}
\usepackage{dcolumn}
\usepackage{graphicx}
\usepackage{epsfig}

\begin{document}

\title{Reply to Comment on "A local realist model for correlations of the singlet state"}

\titlerunning{Reply to Comment on "A local realist model for correlations of the singlet state"}
\authorrunning{H. De Raedt et al}
\author{H. De Raedt\inst{1}, K. Michielsen\inst{2}, S. Miyashita\inst{3}, and K. Keimpema\inst{1}
}                     
%
%
\institute{
Department of Applied Physics,
Zernike Institute for Advanced Materials, University of Groningen, Nijenborgh 4,
NL-9747 AG Groningen, The Netherlands
  \and
EMBD, Vlasakker 21, B-2160 Wommelgem, Belgium
  \and
Department of Physics, Graduate School of Science, University of
Tokyo, Bunkyo-ku, Tokyo 113-0033, Japan
}
\date{Accepted: \today\\Published online: ??}
%
\abstract{
The general conclusion of Seevinck and Larsson is that our model
exploits the so-called coincidence-time loophole and produces sinusoidal (quantum-like)
correlations but does not model the singlet state because it does not violate the relevant Bell inequality
derived by Larsson and Gill,
since in order to obtain the sinusoidal correlations the probability of
coincidences in our model goes to zero.
In this reply, we refute their arguments that lead to this conclusion
and demonstrate that our model can reproduce results of photon and ion-trap experiments
with frequencies of coincidences that are not in conflict with the observations.
} 

\def\url#1{{\tt #1}}

\maketitle

In order to come to their conclusions, Seevinck and Larsson made the following statements~\cite{SEEV07}:

\begin{itemize}
\item{De Raedt {\it et al.} claim that their model violates the CHSH inequality, a claim that cannot be found in Ref.~\cite{RAED06c};}
\item{The CHSH inequality is inappropriate for models that exploit the so-called coincidence-time loophole~\cite{LARS04}
and the appropriately modified inequality~\cite{LARS04} is not violated by the model of De Raedt {\it et al.};}
\item{The model of De Raedt {\it et al.} cannot reproduce all experimental realizations of the EPRB experiment;}
\item{De Raedt {\it et al.} claim that their model can reproduce the coincidences of recent experimental results,
another claim that cannot be found in Ref.~\cite{RAED06c};}
\end{itemize}
and put our model in the context of hidden variable models to obtain an expression for the probability of coincidences.

In this reply, we point out once more that in our work~\cite{RAED06c} we did not rely on Bell's or CHSH's inequality
nor on any generalization thereof to come to our conclusion that it is possible to construct an event-based
simulation model that satisfies Einstein's criteria of local causality and realism and can reproduce the expectation
values of a system of two $S=1/2$ particles in the singlet state. In our work~\cite{RAED06c} we did not make
any claims about these inequalities, neither did we make any statement about coincidences in real experiments.
We furthermore demonstrate that the calculation by Seevinck and Larsson~\cite{SEEV07} of the probability of coincidences
for our model is simply wrong and we present results for the frequency of coincidences which
compare rather well to the values observed in recent experiments.

Before replying in detail to the comments of Seevinck and Larsson~\cite{SEEV07} on our model~\cite{RAED06c},
we first want to sincerely apologize that we did not make a reference to the model presented in Ref.~\cite{LARS04}
which, like our model, uses coincidence in time as a criterion to decide which pairs of detection events are to be
considered as stemming from a single two-particle system.
Furthermore, to our knowledge, it is the first work to point out how the original Bell inequality changes when
using this post-selection procedure~\cite{LARS04}.
Although in our work~\cite{RAED06c}, we did not rely on Bell's
original inequality or on any of its generalizations~\cite{BELL93} to come to our conclusions, we
should have made reference to Ref.~\cite{LARS04} only because of the fact that the model presented in Ref.~\cite{LARS04} uses the
same pair selection criterion as we use in our model~\cite{RAED06c}.

As stated in paragraph 2 of our paper~\cite{RAED06c},
we consider the original EPRB problem, that is the construction
of a model that satisfies Einstein's criterion of local causality
for each pair of events and reproduces the expectation values
of a system of two $S=1/2$ particles in the singlet state.
Whether or not this model leads to a violation
of some inequality is of secondary interest.

We do not share the point of view of Seevinck and Larsson~\cite{SEEV07}
that a system is in the "singlet state" if and only if
some correlation violates a certain bound and if the
probability of coincidences does not go to zero.
This viewpoint does not make much sense if we apply it to the ground state
of a hydrogen molecule (a spin singlet), for instance.
The singlet state is a concept of quantum theory.
Unlike Seevinck and Larsson suggest in their conclusion, the singlet
is not defined with any reference to coincidence counts
or (generalized) Bell inequalities.

According to quantum theory, the singlet state is completely characterized by the single-particle
and two-particle expectation values $E_1({\bf a}_1)=E_2({\bf a}_2)=0$ and
$E({\bf a}_1,{\bf a}_2)=-{\bf a}_1\cdot {\bf a}_2$, respectively. Because quantum theory has nothing to say
about single events~\cite{HOME97}, it does not give us a recipe to compute $E_1({\bf a}_1)$, $E_2({\bf a}_2)$ and
$E({\bf a}_1,{\bf a}_2)$ from the record of single events in laboratory experiments or theoretical models.

In the case of EPRB laboratory experiments with photons~\cite{WEIH98}, coincidence in time seems
to be a good criterion to identify particle pairs based on the time-tag data of single particle events,
since by using this criterion results comparable to those expected from quantum theory can be obtained.
Therefore we use the same criterion in our model~\cite{RAED06c}. However, other criteria to decide
which single particles belong to a single two-particle system are not excluded.
The criterion depends on the experimental setup but quantum theory does not give any guidance to
define a criterion.
Having made a definite choice for this criterion,
we can compute the single-particle and two-particle expectation values from
the record of single events in laboratory or computer experiments and
compare the outcome with the results from quantum theory.
If and only if we find $E_1({\bf a}_1)=E_2({\bf a}_2)=0$ and
$E({\bf a}_1,{\bf a}_2)=-{\bf a}_1\cdot {\bf a}_2$,
we may say that we found expectation values that correspond to those of a singlet state.
No other criteria, like violating an inequality or computing a probability of
coincidences for example, are required to come to this conclusion.
Note that we cannot say anything more than that the
expectation values correspond to those of a singlet state.
For example, we cannot make statements such as the source produces singlets,
since the results for the expectation values do not only
depend on the characteristics of the source and the detection elements
but also on the measurement (post-processing) process.

It is self-evident that our model is too simple to describe, in every detail,
all conceivable experimental realizations of the EPRB thought experiment but
it is the first model that satisfies Einstein's conditions of local
causality and realism and that exactly reproduces the single-particle and two-particle expectation values of the singlet state.
In this respect, it may be viewed as the first realization of the EPRB thought experiment (as defined by EPRB),
since none of the laboratory experiments of the EPRB experiment have shown results
for the single- and two-particle expectation values that compare so well with those of quantum theory.
In those experiments, conclusions are usually drawn based on the value for $S_{max}$ only.
Moreover, to draw conclusions about local realist modelling of expectation values
that agree with those of a singlet state, finding one such model is sufficient.
Whether this model then fails to describe all possible laboratory realizations of the EPRB thought experiment
becomes irrelevant and it remains to be seen if these laboratory experiments
produce data that completely characterize a singlet state, a requirement of
the EPRB thought experiment.

Seevinck and Larsson state ``we will put the model used by De Raedt {\sl et al.} in its proper context''\cite{SEEV07}.
However, they failed to do so in any respect.
In spite of the fact that in our paper, we repeatedly stress that in formulating
our model we do not rely on concepts of probability theory, they seem to ignore our statements.
This is most evident by their statement that "The local hidden variable ...
is denoted by $S_{n,i}$".
Since, according to Larsson a hidden-variable model is really a probabilistic model~\cite{LARS98}
and since our model is purely ontological, the concept of a hidden variable cannot be applied to our model as such.

In contrast to the (repeated) statement made in Ref.~\cite{SEEV07}, we did not
claim that the CHSH inequality (see Eq.~(2) in Ref.~\cite{SEEV07}) is valid for our model.
There is no such statement in our paper.
In our paper, we studied the values of $S_{max}$ as a function of the time window $W$ relative
to the time-tag resolution $\tau$ and this for several values of the model parameter $d$. We compared the results
with $S_{max}=2\sqrt{2}$, the quantum theoretical result for the singlet state and also the
maximum value for $S_{max}$ that can be obtained for any choice of the quantum state. Although we find that for some
model parameters $2<S_{max}<4$ we did not claim that our model violates the CHSH inequality,
as stated by Seevinck and Larsson~\cite{SEEV07}.
In fact, using elementary algebra it follows immediately from Eqs.~(3) and (5) of Ref.~\cite{RAED06c} that
$|E({\bf a}_1,{\bf a}_2)|\le1$ and that
\begin{equation}
\label{eq400}
|E({\bf a} ,{\bf c} )-E({\bf a},{{\bf d} })+E({{\bf b} },{\bf c} )+E({{\bf  b} },{{\bf d} })|\le4,
\end{equation}
for the data generated by our computer model.
Without any further constraints on the algorithm that generates the
data $\{\Upsilon_1,\Upsilon_2\}$ (see Eq.~(1) in Ref.~\cite{RAED06c}),
the upperbound (4) in Eq.~(\ref{eq400}) cannot be improved.
In our paper~\cite{RAED06c}, we use expression Eq.~(5) (see Ref.~\cite{RAED06c}) to discuss the nature of the quantum state,
but attach no meaning to the violation of some bound by our simulation data.

Not surprisingly, also the statement "They furthermore claim that the maximal quantum violation is ...",~\cite{SEEV07}
is wrong. We did not make any reference to the CHSH inequality in our paper.
Looking at Fig.~3 of our paper~\cite{RAED06c}, Seevinck and Larsson should have noted that for $d>3$, our model
can produce correlations that are (much) stronger than those of quantum theory
(which in view of Eq.~(\ref{eq400}) is not a surprise).
In fact, the correct statement (see Ref.~\cite{RAED06c}) is that our model can exhibit correlations that
are {\bf stronger} than those of quantum theory of two $S=1/2$ particles.

As we mentioned before, we agree with Seevinck and Larsson that when time-coincidence is used to decide
which pairs of detection events are to be considered as stemming from a single two-particle system
and if one would like to consider a generalized Bell inequality,
the relevant inequality to consider would be Eq.(4) in Ref.~\cite{SEEV07} and not the CHSH inequality.
In this modified inequality $\gamma$ is the infimum of the probability of coincidence~\cite{LARS04}.
Seevinck and Larsson compute $\gamma$ for our model. They conclude that our model does not violate Eq.(4) in
Ref.~\cite{SEEV07}.
Moreover, although in the paper on the photon experiment~\cite{WEIH98}
or in the paper on the ion-trap experiment~\cite{ROWE01}, there is no information
about the minimum frequency of coincidences (the minimum is required for the application of
Eq.(4) in Ref.~\cite{SEEV07}), Seevinck and Larsson refer to these papers
when they cite the values of $\gamma=0.05$ and $\gamma=1$ and then state that
our model cannot reproduce the frequencies of coincidences that agree with
those found in these two experiments.
First of all, the statement that ``De Raedt {\sl et al.} claim that their model can reproduce the coincidences
of recent experimental results''~\cite{SEEV07} is simply wrong: There is no such claim in our paper~\cite{RAED06c}.
Second, it is logically inconsistent to draw conclusions based on the comparison with the ion-trap experiment for
which $\gamma=1$~\cite{SEEV07} and third, the calculation of $\gamma$~\cite{SEEV07} for our model is wrong, two statements
which we will prove in what follows.

Let us formalize the statements in Ref.~\cite{SEEV07} as propositions (denoted by A,B,...):
\begin{enumerate}
\item[A.]{The probabilistic, hidden variable model using the time window to define coincidences~\cite{LARS04}
yields the inequality given in Eq.(4) of Ref.~\cite{SEEV07} (see also Ref.~\cite{LARS04}). The upperbound of this
inequality is given by $6/\gamma -4$, where $\gamma$ denotes the
infimum of the probability of coincidences over all possible settings ${\bf a}_1,{\bf a}_2$ of the detectors.}
\item[B.]{In the EPRB experiment with ions~\cite{ROWE01}, $\gamma=1$~\cite{SEEV07}.}
\item[C.]{The probabilistic, hidden variable model of Ref.~\cite{LARS04} applies
to the ion-trap experiment~\cite{ROWE01} and hence the experimental data
should satisfy the inequality given in Eq.(4) of Ref.~\cite{SEEV07}.
Note that the second part of this statement implicitly follows from proposition B.}
\item[D.]{The ion-trap experiment yields $S_{max}\approx2.25$~\cite{ROWE01}.
Strictly speaking, this statement is not made in Ref.~\cite{SEEV07},
but it is an experimental fact and as such cannot be denied.}
\end{enumerate}
Let us now apply the rules of elementary logic.

If $\gamma=1$~\cite{SEEV07}, the ion-trap experiment not only violates the original Bell inequality
but as $6/\gamma-4=2$, it also violates the inequality given in Eq.(4) of Ref.~\cite{SEEV07}.
Thus, we have
\begin{equation}
\label{eq43}
\hbox{A}\wedge \hbox{B}\wedge \hbox{C}\wedge \hbox{D}\Rightarrow \overline{C},
\end{equation}
where $\wedge$, $\Rightarrow$ and $\overline{\phantom{C}}$ denote the logical ``and'' operation,
logical implication, and logical negation, respectively.
Clearly, Eq.~(\ref{eq43}) expresses a logic contradiction.
If we assume that propositions A and D are true (as we do),
then we must conclude that B or C or both B and C are false.
In any case, the argument used by Seevinck and Larsson
leads to a logical contradiction, independent of what
we wrote in our paper~\cite{RAED06c}.

We should not exclude the possibility (that is, we might
accept proposition $\overline{\hbox{C}}$) that
the model of Ref.~\cite{LARS04} or ours~\cite{RAED06c}
is too simple to describe the ion-trap experiment~\cite{ROWE01}.
This experiment uses a detection pulse during which the bright state of an ion
scatters many photons (64 on average)~\cite{ROWE01}.
This process may not be sampling ``single-events'' but is
more likely to probe the ensemble average that is given by quantum theory
(although the number of samples, $\approx64$, is not large).

By trying to put our work in the context of "hidden variable theories",
Seevinck and Larsson also made mistakes in elementary algebra.
Seevinck and Larsson assume that the probability of coincidences is given by the denominator of Eq.~(6) in
Ref.~\cite{RAED06c} (see Appendix A of Ref.~\cite{SEEV07}),
from which they derive an expression for the probability of coincidences $\gamma$ (see Eq.~(8)
in Ref.~\cite{SEEV07}).
However, Seevinck and Larsson apparently overlooked the fact that in going from Eq.~(3) to Eq.~(6) (see Ref.~\cite{RAED06c}),
we take the limit $W/T_0=\tau/T_0\rightarrow0$ and
let the number of events $N$ in both the numerator and denominator go to infinity.
Although the ratio remains finite, which is obvious in the case ${\bf a}_1={\bf a}_2$ ($x_1x_2=-1$)
where it is equal to minus one, the limit of the denominator may not exist and in fact,
it diverges if ${\bf a}_1={\bf a}_2$.
This is not a problem of our model: This divergence merely
signals that one has to be careful in taking the limits.
The mathematical derivation in Appendix A of Ref.~\cite{SEEV07} is simply incorrect.

Nevertheless, Seevinck and Larsson raise an interesting question about the role of the
frequency (not probability) of coincidences in our model.
For nonzero time-tag resolution $\tau$ and time window $W\ge\tau$,
the frequency of coincidences in our simulation model is given by
\begin{equation}
\label{eq402}
\Gamma=\frac{1}{N}\sum_{n=1}^{N}\Theta(W-\vert t_{n,1} -t_{n ,2}\vert)
,
\end{equation}
a well-defined quantity in our simulation model that is easy to compute numerically.
Notice that $\Gamma$ is a function of ${\bf a}_1$ and ${\bf a}_2$
and that $0\le\Gamma\le1$.
Assuming that the results we obtain by using pseudo-random numbers
can be described by a probabilistic model,
we expect that $\gamma=\min_{{\bf a}_1,{\bf a}_2}\Gamma$
with probability one if $N$ is sufficiently large.
With these additional assumptions,
not only the inequality Eq.~(\ref{eq400}) holds but
also the inequality given by Eq.~(4) of Ref.~\cite{SEEV07} holds.

In our model, there are four free parameters, namely
the time window $W/\tau$, the maximum time delay $T_0/\tau$,
the time-tag exponent $d$ and the number of events $N$.
For $d=3$, $N\rightarrow\infty$ and in the limit $W/T_0=\tau/T_0\rightarrow0$,
our model reproduces exactly, the expression for the two-particle expectation value of a quantum system
in the singlet state~\cite{RAED06c}.

For $d=3$, $W=\tau$, $T_0/\tau=1000$ and $N=10^6$
(the results reported in this paper do not change if $N>5\times10^5$),
we find that $\min_{{\bf a}_1,{\bf a}_2}\Gamma\approx1.27W/T_0$,
in concert with the rigorous result (for $d=3$ and $W=\tau$)
\begin{equation}
\label{eq403}
\min_{{\bf a}_1,{\bf a}_2}\left\{\lim_{W/T_0\rightarrow0}\Gamma\right\}=\frac{4}{\pi}\frac{W}{T_0}\approx1.27\frac{W}{T_0}
.
\end{equation}
Thus, in the regime $W/T_0=\tau/T_0\rightarrow0$, we find
that the minimum frequency of coincidences is proportional to the width
of the time bins, as it should be.

Next, we consider the possibility of fitting the results of our model to
the experimental data of an EPRB experiment with photons~\cite{WEIH98} and ions~\cite{ROWE01}.
According to Seevinck and Larsson~\cite{SEEV07}, our model cannot reproduce these experimental data.
For simplicity, we set $T_0/\tau=1000$, $d=3$ and take $N=10^{6}$.
Then, there is one free parameter left, namely the (dimensionless) time-window $W/\tau\ge1$.
The fitting procedure consists of changing $W/\tau$ such that
the value of $S_{max}=\max_{{\bf a},{\bf b},{\bf c},{\bf d}}S({\bf a},{\bf b},{\bf c},{\bf d})$
agrees with values cited in Refs.~\cite{WEIH98,ROWE01}.

\begin{figure}[t]
\begin{center}
\mbox{
\includegraphics[width=8cm]{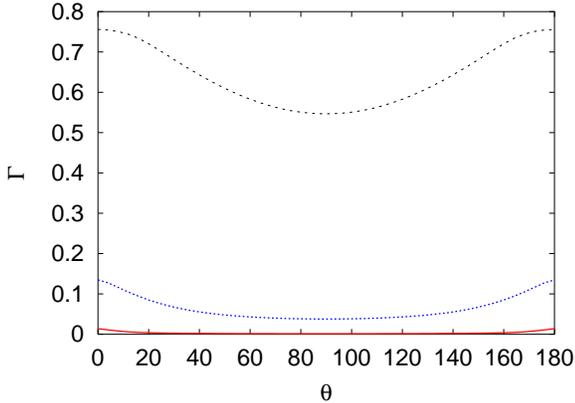}
}
\caption{(color online)
The frequency of coincidences $\Gamma$ as a function of $\theta=\arccos({\bf a}_1\cdot{\bf a}_2)$
for parameters $\tau/T_0$, $W/T_0$ and $d$ chosen such (see text)
that the simulation model reproduces the result for a singlet state,
$S_{max}=2.83$ (solid line, red), and the values of
$S_{max}=2.25$ (dashed line, black) and
$S_{max}=2.73$ (dotted line, blue),
as obtained from experiments
with ions~\cite{ROWE01}
and
with photons~\cite{WEIH98}, respectively.
}
\label{gamma}
\end{center}
\end{figure}

In Fig.~\ref{gamma}, we present our results for the frequency
of coincides for the values of $W/\tau=285$, $W/\tau=16$, and
$W/\tau=1$, for which our model yields
$S_{max}=2.25$~\cite{ROWE01},
$S_{max}=2.73$~\cite{WEIH98},
and
$S_{max}=2.83$ (singlet state), respectively.

From Fig.~\ref{gamma}, we see that in order to reproduce the ion-trap result~\cite{ROWE01},
the frequency of coincidences $\Gamma\ge0.52$ is quite large.
It is important to recognize that with four free parameters at our disposal,
it is easy to reproduce almost any number for $S_{max}$, as long as it is between zero and four.
For instance, we find the same value of $S_{max}=2.25$ for
$W=\tau$ and $T_0/\tau=1.025$ but then $\Gamma\ge0.87$.
In any case, these results refute the statement in the Comment that our model
cannot reproduce the experimental result $S_{max}=2.25$
of the ion-trap experiment~\cite{ROWE01} with a nonzero value of $\Gamma$.
Fitting our model (for $d=3$ and $T_0/\tau=1000$) to $S_{max}=2.73$ and $S_{max}=2.83$
yields $\Gamma>0.0377$ and $\Gamma>0.00127$, respectively.

An analysis of experimental data for an EPRB experiment with photons~\cite{WEIHdownloadShort}
yields $\Gamma\approx0.01$
(the value of $\gamma\approx0.05$ cited
in Ref.~\cite{SEEV07} is the total frequency of coincidences, that is the sum over four experiments,
and not the infimum over all possible experiments, as required for the application
of the inequality given in Eq.~(4) of Ref.~\cite{SEEV07}).
Thus, for the same value of $S_{max}$, our model
yields a value of $\Gamma$ that is larger ($\Gamma=0.0377$)
than the value that can be extracted from
experimental data for an EPRB experiment with photons~\cite{WEIHdownloadShort}.

As our model is flexible enough to yield for $S_{max}$ any number between zero and four
with reasonable values of the model parameters,
it is of interest to study how the correlation $E({\bf a}_1,{\bf a}_2)$ deviates from the
result $E({\bf a}_1,{\bf a}_2)=-{\bf a}_1\cdot {\bf a}_2$ of a system in the singlet state as we fit
the values of $S_{max}$ to the experimental results.

\begin{figure}[t]
\begin{center}
\mbox{
\includegraphics[width=8cm]{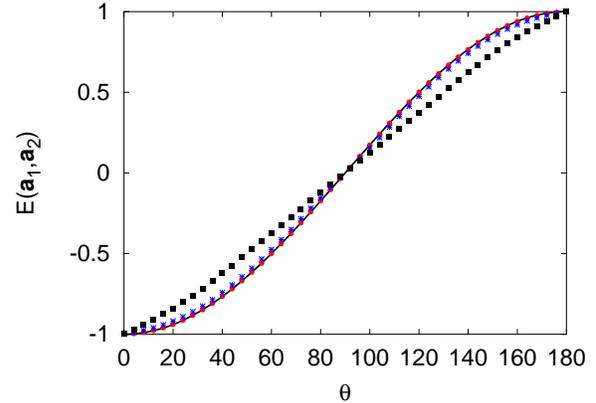}
}
\caption{(color online)
Simulation results of the two-particle correlation $E({\bf a}_1,{\bf a}_2)$ as a function of
$\theta=\arccos({\bf a}_1\cdot{\bf a}_2)$
for the model parameters that yield
$S_{max}=2.25$ (squares, black),
$S_{max}=2.73$ (stars, blue),
and
$S_{max}=2.83$ (bullets, red),
respectively.
The solid line (black) is the result ($E({\bf a}_1,{\bf a}_2)=-{\bf a}_1\cdot{\bf a_2}$) for the singlet state.
}
\label{gamma2}
\end{center}
\end{figure}

In Fig.~\ref{gamma2}, we show the simulation results
for the same three cases $S_{max}=2.25,2.73,2.83$.
From Fig.~\ref{gamma2}, it is clear that the simulation data that yields
$S_{max}=2.25,2.73$ cannot be described by a single sinusoidal function,
but for $S_{max}\ge2.73$ the deviations from a single sinusoidal
are small and it remains to be seen if experiments can resolve such small differences.

As is evident from Fig.~3 in Ref.~\cite{RAED06c},
for $d>3$ our model yields the value for the singlet state $S_{max}=2\sqrt{2}$
without having to consider the limit $W/T_0=\tau/T_0\rightarrow0$.
Thus, in order for an experiment and a model of the type considered in our paper
to reproduce {\bf all} the features of a quantum system
of two $S=1/2$ particles in the singlet state, it is not sufficient to
show that it can yield $S_{max}=2\sqrt{2}$ for some
choice of the parameters. As mentioned before, the singlet state is completely characterized by the
single and two-particle expectation values. Hence, in order to make a comparison with the singlet state,
it is necessary to measure or compute these two quantities.

Finally, the statement in our paper~\cite{RAED06c}
that ``Our work suggests that it is possible to construct event-based simulation models
that satisfies Einstein's criteria of local causality and realism
and can reproduce the expectation values calculated by quantum theory~\cite{RAED05b,RAED05c,RAED05d,MICH05,RAED06a}''
should not be taken out of the context as Seevinck and Larsson did by omitting the references.
In fact, what we have shown in the work that we refer to is that it
is possible to perform an event-based simulation, satisfying Einstein's criteria of local
causality, of a universal quantum computer~\cite{MICH05}, which according to the theory of quantum computation
should suffice to simulate any quantum system~\cite{NIEL00}.

In conclusion, the purpose of our work is to construct an event-based simulation model,
satisfying Einstein's criteria of local causality and realism, that produces the quantum correlations of the singlet
state~\cite{RAED06c}.
As we have shown in Ref.~\cite{RAED06c} we succeeded, to our knowledge for the first time, in constructing such a model.
Our conclusion that we find results that are indistinguishable from those of a singlet state is
based on the fact that the calculated single-particle averages and two-particle correlation function
agree with the well-known results $E_1({\bf a}_1)=E_2({\bf a}_2)=0$ and
$E({\bf a}_1,{\bf a}_2)=-{\bf a}_1\cdot {\bf a}_2$ for a system of two $S=1/2$ particles in the singlet state.

We have also demonstrated in Ref.~\cite{RAED06c} and in this reply that knowing $S_{max}$,
a quantity derived from the two-particle correlation function,
does not suffice to draw any conclusion about the observation of a singlet(-like) state.
We also demonstrate in this reply, that our model can not only produce the results from quantum theory
for a system of two $S=1/2$ particles in the singlet state but that it can also be applied to
EPRB laboratory experiments with photons and ions and give
results for $S_{max}$ and the frequency of coincidences that are comparable
to the values extracted from these experiments.

\bibliographystyle{epj}
\bibliography{../epr}

\begin{thebibliography}{15}

\bibitem{SEEV07}
M.P. Seevinck, J.A. Larsson, Euro. Phys. J. B \textbf{?}, ?  (2007)

\bibitem{RAED06c}
K.~{De Raedt}, K.~Keimpema, H.~{De Raedt}, K.~Michielsen, S.~Miyashita, Euro.
  Phys. J. B \textbf{53}, 139  (2006)

\bibitem{LARS04}
J.A. Larsson, R.D. Gill, Europhys. Lett. \textbf{67}, 707  (2004)

\bibitem{BELL93}
J.S. Bell, \emph{Speakable and unspeakable in quantum mechanics} (Cambridge
  University Press, Cambridge, 1993)

\bibitem{HOME97}
D.~Home, \emph{Conceptual Foundations of Quantum Physics} (Plenum Press, New
  York, 1997)

\bibitem{WEIH98}
G.~Weihs, T.~Jennewein, C.~Simon, H.~Weinfurther, A.~Zeilinger, Phys. Rev.
  Lett. \textbf{81}, 5039  (1998)

\bibitem{LARS98}
J.A. Larsson, Phys. Rev. A \textbf{57}, 3304  (1998)

\bibitem{ROWE01}
M.A. Rowe, D.~Kielpinski, V.~Meyer, C.A. Sackett, W.M. Itano, C.~Monroe, D.J.
  Wineland, Nature \textbf{401}, 791  (2001)

\bibitem{WEIHdownloadShort}
\url{http://www.quantum.at/research/photonentangle/\-bellexp/data.html}.

\bibitem{RAED05b}
K.~{De Raedt}, H.~{De Raedt}, K.~Michielsen, Comp. Phys. Comm. \textbf{171}, 19
   (2005)

\bibitem{RAED05c}
H.~{De Raedt}, K.~{De Raedt}, K.~Michielsen, J. Phys. Soc. Jpn. Suppl.
  \textbf{76}, 16  (2005)

\bibitem{RAED05d}
H.~{De Raedt}, K.~{De Raedt}, K.~Michielsen, Europhys. Lett. \textbf{69}, 861
  (2005)

\bibitem{MICH05}
K.~Michielsen, K.~{De Raedt}, H.~{De Raedt}, J. Comput. Theor. Nanosci.
  \textbf{2}, 227  (2005)

\bibitem{RAED06a}
H.~{De Raedt}, K.~{De Raedt}, K.~Michielsen, S.~Miyashita, Comp. Phys. Comm.
  \textbf{174}, 803  (2006)

\bibitem{NIEL00}
M.~Nielsen, I.~Chuang, \emph{Quantum Computation and Quantum Information}
  (Cambridge University Press, Cambridge, 2000)

\end{thebibliography}

\end{document}